# Diversity and Social Network Structure in Collective Decision Making: Evolutionary Perspectives with Agent-Based Simulations

Shelley D. Dionne,[1,2,3] Hiroki Sayama,[1,2,4,5] and Francis J. Yammarino[1,2,3]

[1] Bernard M. & Ruth R. Bass Center for Leadership Studies
[2] Center for Collective Dynamics of Complex Systems
[3] School of Management
[4] Department of Systems Science and Industrial Engineering
Binghamton University, State University of New York, Binghamton, NY 13902-6000, USA.

[5] School of Commerce, Waseda University, Shinjuku, Tokyo 169-8050, Japan.

Correspondence should be addressed to Hiroki Sayama; sayama@binghamton.edu

## Abstract

Collective, especially group-based, managerial decision making is crucial in organizations. Using an evolutionary theoretic approach to collective decision making, agent-based simulations were conducted to investigate how human collective decision making would be affected by the agents' diversity in problem understanding and/or behavior in discussion, as well as by their social network structure. Simulation results indicated that groups with consistent problem understanding tended to produce higher utility values of ideas and displayed better decision convergence, but only if there was no group-level bias in collective problem understanding. Simulation results also indicated the importance of balance between selection-oriented (i.e., exploitative) and variation-oriented (i.e., explorative) behaviors in discussion to achieve quality final decisions. Expanding the group size and introducing non-trivial social network structure generally improved the quality of ideas at the cost of decision convergence. Simulations with different social network topologies revealed collective decision making on small-world networks with high local clustering tended to achieve highest decision quality more often than on random or scale-free networks. Implications of this evolutionary theory and simulation approach for future managerial research on collective, group, and multi-level decision making are discussed.

## Introduction

Collective decision making plays an increasingly important role in society and organizations today (Mannes 2009, Kerr and Tindale 2004, Dionne et al. 2010, McHugh et al. 2016, Uitdewilligen and Waller 2018, Dionne et al. 2018). In high-tech industries, for example, the number of engineers participating in the design of a single product can amount to hundreds or even thousands due to the increase of the product's complexity far beyond each individual engineer's capacity, which almost inevitably results in suboptimal outcomes (Klein et al. 2003, Braha et al. 2006, ElMaraghy et al. 2012). Another example is the online collective decision making among massive anonymous participants via large-scale computer mediated communication networks, including collective website/product rating and common knowledge base formation (O'Reilly 2005, Economist 2009). In these and related cases, participants and their societal or organizational structure may influence the final outcome of decision making processes. The complexity of the process is more pronounced when the participants are heterogeneous and are embedded in a topologically non-uniform network with differential distribution of power, as in most organizations and social systems (Dionne et al. 2010). The dynamics of human collective decision making in such conditions are poorly understood, and as such pose significant challenges for the social and organizational sciences.

Evidence of these challenges exist within the leadership, psychology and organizational behavior/management disciplines where collective dynamics, using both experimental and applied studies, generally emphasize linear statistical relationships between specific, narrowly defined team- or individual-level variables (Kerr and Tindale 2004, Salas et al. 2004, Dionne et al. 2012, Humphrey and Aime 2014). Traditional studies seldom account for nonlinear dynamical processes that take place in a high-dimensional problem space and/or non-trivial social structure where interactions occur within a networked organizational structure. Abbott (2001) highlights this problem within the social sciences by discussing a "general linear reality," where mainstream social science theories and methods treat linear models as actual representations of social systems.

Examples of recent research not necessarily following a "general linear reality" to model inherent complexity in social systems are found within the complex systems research community, where social processes are studied using a mathematical/computational modeling approach (Bar-Yam 1997, 2004, Braha et al. 2006, Epstein 2006, Miller and Page 2007, Castellano et al. 2009, Dionne et al. 2010, Couzin et al. 2011, Sayama et al. 2011, McHugh et al. 2016, Giannoccaro et al. 2018, Page 2018). Because emphasis is on emergent dynamical behavior of systems caused by nonlinear interactions among massive numbers of parts (a pervasive phenomenon also found in fields such as physics, biology, sociology, psychology, economics, engineering and computer science), advances in modeling complex systems may be applied to benefit organizational research (Carroll and Burton 2000, Schneider and Somers 2006). However, many of these complex systems models were developed in non-human contents such as physics and biology, and thus their model assumptions often would be too simplistic to capture the complexity of collective human decision making.

The aim of this research is to reveal how we may be able to enhance performance of groups and other entities involved in collective human decision making by expanding computational models of social systems to complex problem domains and by applying them to predict the effects of individual and collective variables upon decision making performance. Collective decision making implies a larger clustering of individuals with interdependency based on shared expectations or hierarchy. Collectives can be complicated structures and include individuals, groups, and even much larger social networks (Dansereau et al. 1984, Yammarino et al. 2005, McHugh et al. 2016, Yammarino and Dionne 2018). We seek to improve our understanding of both the dynamic nature of the collective decision process (Waller et al. 2016), as well as the influence of diversity and social connectivity issues related to decision making among a number of participating group members. Our unique contributions include employing evolutionary views in understanding decision making (Sayama and Dionne 2015), which enables a straightforward, mechanistic explanation of many empirical findings about the effects of group composition and dynamics on group performance. Considering specific within-group level issues regarding the collaborative process of decision making also may offer clarity regarding the influence of group composition on performance.

We first explore how evolutionary theory can address complex changes over time by providing an explanatory framework for collective decision processes, and then discuss how specifying a targeted level of analysis can inform appropriate interpretation and limitations of decision making in dynamic environments. Finally, a computational agent-based model (Epstein 2006, Miller and Page 2007, Page 2018) with an evolutionary focus on collective decision making in groups and social networks is developed and tested, with diversity of problem understanding, behavioral patterns and social network structure manipulated as experimental variables. This approach is similar to Kozlowski and colleagues' (2013) recommendations for capturing multilevel dynamics of emergence through development of a conceptual foundation and integration of agent-based modeling as part of a theory testing process.

Specifically, this study adapts four recommendations from Meyer et al. (2005) to advance our theoretical understanding of collective decision making in complex social systems: 1) consider the impact of time by constructing a dynamical simulation model; 2) study situations in flux by situating interacting agents in a continuously changing social environment; 3) incorporate nonlinear concepts by utilizing evolutionary theory that naturally represents nonlinearity in the exploration of a complex problem space; and 4) design multi-level research by taking into account within- and between-group differences as well as complex social network topologies. These guidelines provide a starting point for investigating the complexity of collective decision making with an evolutionary and multi-level, network-oriented framework. Prior dynamical modeling in organizational research may have considered the impact of time and situations in flux; few if any, however, have included specific evolutionary and multi-level, network-oriented concepts.

## Backgrounds

**Evolutionary Theory and Collective Decision Making**

Evolutionary theory describes adaptive changes of populations primarily by combining mechanisms of variation and selection (Futuyma 2005, Wilson 2005). The roles of these two mechanisms are similar to "exploration" and "exploitation" in organizational learning literature (Cheng and Van de Ven 1996, He and Wong 2004, March 1991). In biological evolution, variation is caused primarily by internal genetic mechanisms (e.g., mutation and recombination) and plays an exploratory role that could potentially lead to a novel possibility of life form, but it usually reduces immediate competitiveness of a population. In contrast, selection is caused primarily by external environment (e.g., natural and sexual selection) and plays an exploitative role that enhances the presence of successful entities (genes, individuals, or groups) and eliminates unsuccessful ones, reducing the number of possibilities while potentially improving the overall competitiveness of the population. A dynamically maintained balance of the two mechanisms is the key to a successful evolutionary adaptation (Mitchell et al. 1991).

We propose human decision making processes within a collective (such as a group or an organization) also may be viewed through a similar lens, by shifting the viewpoint from individual members' personal properties (a more traditional psychological and decision making approach) to dynamical changes of ideas being discussed within the collective, where populations of potential ideas evolve via repetitive operations such as reproduction, recombination, mutation, and selection of ideas, conducted by participating human individual members acting as the environment for the ideas (Sayama and Dionne 2015). Table 1 provides a summary of the evolutionary framework we propose by illustrating how some key evolutionary theoretic concepts can be linked to components of human decision making processes. We take this approach because evolutionary theory provides a powerful theoretical framework that can readily address complex changes of systems over time in extremely high-dimensional problem space, while its explanatory mechanisms (heredity, variation, and selection) are theoretically clean-cut and easily accessible (Wilson 2005). Moreover, by shifting the viewpoint from individuals to ideas, a model could be liberated from the commonly used but somewhat artificial assumption that each individual always has his/her decision in mind. Rather, various ideas developed within and among participants are collectively reflected in the idea population, to which diverse within-individual cognitive/behavioral patterns can be easily applied as a set of multiple evolutionary operators simultaneously acting on the same, shared idea population. Shifting a viewpoint away from individuals has precedence in event-level literatures as well (Hoffman and Lord 2013, Morgeson et al. 2015).

## TABLE 1

**Evolutionary Concepts Applied to Corresponding Decision Making Process Components**

| Evolutionary Concept | Decision Making Component |
|---|---|
| Genetic possibility space | Problem space (decision space) |
| Genome | Potential idea (a set of choices for all aspects of the problem) |
| Locus on a genome | Aspect of the problem |
| Allele (specific gene) on a locus | Specific choice made for an aspect |
| Population | A set of potential ideas being discussed |
| Fitness | Utility value of a potential idea (either perceived or real) |
| Adaptation | Increase of utility values achieved by an idea population |
| Selection | Narrowing of diversity of ideas based on their fitness |
| Replication | Increase of relative popularity of a potential idea in the discussion |
| Recombination | Production of a new potential idea by crossing multiple ideas |
| Mutation | Point-like change in an idea (possibly coming up with a novel idea); can be *random* (unpremeditated change) or *intelligent* (premeditated change) |

Note: Adapted from Sayama and Dionne (2015)

**Evolutionary Operators and Collective Decision Processes**

Various human behaviors in discussion and decision making processes may be mapped to several evolutionary operators (Mitchell 1996, Sayama and Dionne 2015; also see Table 1). For example, advocacy of a particular idea under discussion can be considered the replication of an idea, a form of positive selection, where the popularity of an idea is increased within the population of ideas. Another example is criticism against an idea. Giving a critical comment on an idea can be considered a form of negative, subtractive selection, which reduces the popularity of the criticized idea within the population of ideas. These positive and negative forms of selection narrow decision possibilities based on utilities ("fitness") of ideas perceived by participants. Other human behaviors can be understood as more variation-oriented evolutionary operators. For example, asking "what if"-type random questions corresponds to random point mutation in evolution, which makes random minor changes to existing ideas. However, such mutations may occur in a non-random, more elaborate manner in human decision making. Humans can mentally explore several different possibilities, assessing different "what-if" scenarios, and then share the idea with the highest perceived utility. This can be considered an intelligent, or hill-climbing, point mutation (Klein et al. 2003) in the evolutionary framework (which is not present in real biological evolution). Finally, the creation of a new idea by crossing multiple existing ideas can be considered a recombination of genomes in the evolutionary framework. These variation-oriented evolutionary operators promote exploration of various possibilities, potentially at the cost of the utilities (fitness) of ideas.

As summarized in Table 1, we define collective decision making as an evolution of ecologies of ideas. Participating individuals in the collective decision process have populations of ideas that evolve via continual applications of evolutionary operators such as reproduction, recombination, mutation, selection, and migration of ideas. This definition can naturally be extended to a social network setting (Sayama and Dionne, 2015), in which social ties between humans are pathways through which ideas migrate. Thus, there appears to be an intuitive parallel between an evolutionary framework and a collective decision process. Applying an evolutionary theory to collective decision making seems consistent with the spirit of the Meyer et al. (2005) suggestions regarding improvement of research techniques to better reflect situations in flux and nonlinear concepts within an evolutionary framework.

**Levels of Analysis and Evolution**

Evolutionary biologists Wilson and Wilson (2008) reiterate the link between adaptation and a specific regard for levels of analysis in reviewing the history of multi-level selection theory. Their evolutionary perspective on multi-level selection challenges researchers to evaluate the balance between levels of selection, specifically where within-group selection is opposed by between-group selection. This deeper view of a multi-level evolutionary process can be applied to organizational research as well (Yammarino and Dansereau 2011). Research on both levels of analysis within organizational behavior (Dansereau et al. 1984, Klein et al. 1994, Dionne et al. 2014) and on group collaborative processes (Chang and Harrington 2005, 2007, van Ginkel and van Knippenberg 2008, Yammarino et al. 2012) highlight the importance and value of explicitly viewing the heterogeneity and/or homogeneity of the group and/or collective. This homogeneity and heterogeneity perspective can be viewed as a within-level examination, where the entity of interest remains the group, but there can be at least two valid views at the collective level: homogeneity or whole groups (what evolutionary theory refers to as a between-group focus) and heterogeneity or group parts (what evolutionary theory refers to as a within-group focus) (Dansereau et al. 1984, Klein et al.1994, Yammarino et al. 2005, Yammarino and Dansereau 2011). Note that, in both views, we consider the groups in a collective decision making context in which individual participants collaborate toward a shared goal, there is a dependency among them, and therefore, the heterogeneity or group parts view is quite different from studying a mere collection of different individuals that do not form a collaborative group.

The concept of differing perspectives on an entity can provide more specific insights regarding group processes, in that phenomena of interest may be more relevant when groups are homogeneous regarding their membership, but differ in characteristics from other groups. In this wholes condition, all members within a group possess the same (or at least very similar) characteristic, while in the next group all members possess some other characteristics that first group perhaps did not. Another view can be taken concerning amounts of a characteristic present, where members of a group would possess the same amount of a characteristic, while members of the next group also would possess the same characteristic, but all members would have more of that characteristic, or all members would have less of that characteristic.

From a contrasting perspective, phenomena of interest may be more relevant when groups are heterogeneous regarding their memberships. In this case, members within a group would have varying degrees of a characteristic, and the next group also would have members with varying degrees of a characteristic, and the same applies for all groups.

**Decision Research and Levels of Analysis**

Precedent for a broadly applicable modeling approach has been established in the evolving architecture of problem-solving networks (Chang and Harrington 2007). This research enabled consideration of a generic problem-solving environment and assessment of emergence regularity of connectors within the problem environment. Moreover, Chang and Harrington's research related to the modeling of both homogenous agents (2005) and heterogeneous agents (2007) is of interest to our work. Specifically, we use homogeneity and heterogeneity of groups as means for examining levels of analysis issues related to collective and/or group processes.

Although Chang and Harrington's (2005, 2007) modeling examines a more multi-level relationship between agents (individuals) and the larger environment, we are concerned with examining a within-group, collective or collaborative decision process, where individuals would not be considered outside of the group. Our examination of a unique within-level evolutionary process, employing both within-group and between-group perspectives, is a novel view of collaborative decision making and advances the understanding of a collective environment.

A critical distinction of our research is that we are interested in examining a type of process occurring within the group over time, not necessarily the specific variables within the process. Dansereau, Yammarino and Kholes (1999) highlighted the nature of such research on differing perspectives of an entity and entity changes rather than on changes in specific variables over time. Because we are interested in the type of process occurring within the group during decision making, we agree with Dansereau and colleagues (1999) that the variables that characterize the level may change or remain stable, but the level of interest remains the same (in our case, the level of interest remains the group).

Related, diversity and/or homogeneity and heterogeneity of groups and information sharing (Gigone and Hastie 1993, Grand et al. 2016, Stasser and Stewart 1992, Uitdewilligen and Waller, 2018) present an additional layer to the decision process that requires consideration. Nijstad and Kaps (2007) noted that homogeneity of preferences leads to a lack of sharing of unique information within a group, whereas preference diversity prevented premature consensus of the group and facilitated unbiased discussions of preferences. Lightle, Kagel and Arkes (2009) indicated individual heterogeneity in information recall may play a role in failure to identify hidden profiles within groups. Similarly, van Ginkel and van Knippenberg (2008) found that groups in decision tasks performed better when task representations emphasized information elaboration and the group acknowledged they shared the view of the task representation. These findings reinforced that groups tend to focus on finding common ground and reaching consensus, but highlighted the importance of understanding, as a group, the task representation. This shared understanding could be critical to group success and adaptation, and as such, we include an indicator of how well group members share a view of what constitutes the problem.

Although advancements in decision research continue, many continue to focus on individual-level aspects related to a decision maker, such as how they adopt practical behavior rules (Maldonato 2007) or identification of performance moderator functions that may affect individual behaviors in simulated environments (Silverman et al. 2006). While multi-level implications exist in recent decision research (Kennedy and McComb 2014, Nijstad and Kaps 2007, van Ginkel and van Knippenberg 2008), there is limited focus on within-group level aspects of a decision process. Moreover, Maldonato (2007) notes there is likely no best way to view the decision process. As such, there may be some benefit to development of a preliminary model exploring the effect of membership similarity and differences on group-based decision processes from evolutionary and levels of analysis-based perspectives. Development of such a model advances understanding of collective decision making in that it builds on prior key decision research (Chang and Harrington 2007, Kock 2004, Knudsen and Levinthal 2007, Nijstad and Kaps 2007), incorporates the suggestions of improving organizational research offered by Meyer et al. (2005), and incrementally increases the complexity yet fuller understanding of the phenomena represented in prior collective decision models.

**Modeling Dynamic Collective Decision Making**

Building from the above notions, the application of computational modeling to dynamical processes such as collective decision making may enable organizational researchers to more appropriately represent the potential nonlinearity of a collective process. For example, interdisciplinary exchange may have informed recent organizational research which includes several dynamical models proposed for collective decision making over social networks that consist of many interacting individuals (Battiston et al. 2003, Klein et al. 2003). These models, primarily an extension of models developed in theoretical physics, provide a novel, promising direction for research on group dynamics and collective decision making. A limitation of this research and more specifically its ability to model complex social systems, however, is the consideration of only simple problem spaces, typically made of binary or continuous numerical choices between 0 and 1.

Increasingly complex nonlinear problem space has been modeled (Klein et al. 2003, Klein et al. 2006, Rusmevichientong and Van Roy 2003) to consider interdependent networks of multiple aspects of a complex problem. This research, however, was not modeled in a collective, non-trivial societal context. This is not surprising because problems arise with collective decision models in that they commonly assume every individual agent has or makes his/her own decision. Following these assumptions, the collective decision making dynamic is represented as a process of propagation, interaction and modification of individual decisions. This is an over-simplified assumption compared to actual cognition processes and behavior of individuals and collectives (Lipshitz et al. 2001, Salas and Klein 2001). Individuals often keep multiple ideas in mind and may remain undecided during or even after a collective-level decision emerges. The collective decision forms not just through the interactions of individual decisions but also through the more active, dynamic exchanges of incomplete ideas and mental models being developed by every individual (Dionne et al. 2010). Such within-individual mental and behavioral complexity has begun to be included in computational models (c.f., Dionne and Dionne 2008, Knudsen and Levinthal 2007), and should be taken into account to a greater extent in order to investigate the complexity of collective human decision making.

## Methods

In view of the contexts for computational models of social and organizational sciences reviewed in the prior section, we had previously proposed a prototype agent-based model that applied the evolutionary framework introduced above to model collective decision making processes within a small-sized, well-connected social network structure (Sayama and Dionne 2015). This model was used to conduct a specific within-level analysis on how homogeneity or heterogeneity of goals and decision utility functions among participants affect dynamics and the final outcomes of their collective decision making, and the predictions made by this model were also confirmed by human-subject experiments (Sayama and Dionne, 2015). This model was still quite limited, however, since the size of the collective remained small, the agents' evolutionary behaviors were designed in a rather unsystematic, ad hoc manner, and the effect of social network structure was not taken into account. In this sense, it was not developed enough to provide sufficient answers to key research questions on diversity and network structure of the collective as discussed in the previous section.

In this paper, we present a new agent-based model that can directly address those key research questions by implementing a systematic control of agents' behavioral balance between selection-oriented and variation-oriented operators, together with much larger, non-trivial social network structure on which agents exchange ideas locally. In our model, agents collaboratively work on an abstract utility maximization task, without explicit knowledge of the entire structure of their utility functions. Agents may have similar utility functions within the group, but across groups there may exist different utility functions. Such a homogeneous condition can represent a "group wholes" view, in which all members of each particular group share a strong degree of similarity with their groups' unique utility function. Conversely, agents may have different utility functions within the group. Such a heterogeneous condition can represent a "group parts" view, in which unique and/or diverse utility functions prevail within each group, but across groups, this pattern is not unique, as group after group exhibits this same type of uniqueness among its members.

We believe that our approach to social dynamics research can move the social sciences away from an oversimplified view in that it investigates nonlinear change in organizational research (Meyer et al. 2005). Moreover, examining a new theoretical framework is consistent with development of computational models, as Adner et al. (2009) recognize that simulation is generally an exercise in theory building.

**Model Assumptions**

*Groups or social networks.* Our model assumes that $N$ agents are connected to a finite number of other agents via links through which ideas are exchanged. Each agent can memorize or hold multiple ideas in its mind. Multiple copies of a single idea may be present, which represents a form of relative popularity for that idea to the agent. Each agent is initialized with a small number of randomly generated ideas in its mind at the beginning of a simulation. The agents begin to perform a set of actions on the population of ideas in their minds repeatedly for a fixed number of iterations. The order by which the agents take actions is randomized every time, but it is guaranteed that every agent does take exactly one action per iteration. This round-robin format is commonly used in idea sharing phases with decision making techniques such as a nominal group technique and various brainstorm initiatives (Paulus and Yang 2000, Van de Ven and Delbecq 1974). As such, the number of actions performed in a simulation is a product of the number of agents $N$ and the number of iterations $T$.

While other group decision research has modeled hierarchical teams in decision models (c.f., Dionne and Dionne 2008), we make no assumptions regarding predetermined leadership and/or abilities within the team as several teams in organizations are self-led and share leadership responsibilities (Salas and Klein 2001). We investigate the potential impact of varying membership within the group (i.e., no assumption of identical abilities or uniform connectivities in general) on the potential pool of ideas. Since no single person is powerful enough to eliminate an idea from the group (i.e., shared leadership), we assumed that actions were performed on single copies of an idea, not the equivalence set of all idea replicates (described in detail below).

*Utility functions.* The use of utility functions in collective decision research is a natural outgrowth of earlier research by Hollenbeck et al. (1995) noting team decision making theory can be considered an adaptation of individual decision models and decision alternatives can vary along a univariate continuum. This view supplies a multi-level (e.g., group parts and group wholes) perspective and allows for adaptation of individual utility functions throughout a collective decision process. Both factors can be represented and/or captured by collective decision computer models (c.f., Dionne and Dionne 2008). As such, the use of utility functions contributes to the development of this model as well.

We use a similar model setting for the problem space and the utility functions as proposed by Sayama and Dionne (2015). The problem space is defined as an $M$-dimensional binary space, within which there are a total of $2^M$ possible ideas. For a simulation, each possible idea has an inherent utility value given by a *true* utility function $U_T$. None of the agents has direct access to the true utility function. Instead, individual agents perceive idea utility values based on their own individual utility functions $U_j$ constructed by adding noise to the *master* utility function $U_M$. The master utility function $U_M$ may or may not be the same as $U_T$, depending on the possibility of group-level bias (explained below). This initialization reflects the notion that today's organizational problems are too complex for a single individual to solve (i.e., true utility value not available to any of group members), and therefore groups or collectives are assembled to solve problems and make decisions (Klein et al. 2003, Salas and Klein 2001). Ideally, collectives function by bringing unique information from members (i.e., individual utility functions) together in such a way as to produce ideas that exceed an individual's idea development capability (Kerr and Tindale 2004).

Utility values are assigned to every point (idea) in the problem space as follows: First, $n$ random bit strings (zeros and ones) $S = \{v_i\}$ ($i = 1...n$) are generated as representative ideas, where each $v_i$ represents one idea that consists of $M$ bits. One of those generated ideas is assigned the maximum utility value, 1, and another is assigned the minimum utility value, 0. Each of the remaining $n - 2$ ideas is assigned a random number sampled from a uniform probability distribution between 0 and 1. This method guarantees that the entire range of utility values is always from 0 to 1, which makes it easier to compare different simulation results. The detailed shape of the distribution varies within this range for different simulation runs.

The utility values of other possible ideas not included in the representative idea set *S* are defined by interpolation. Specifically, the utility value of an idea *v* not present in *S* is calculated as a weighted average of the utility values of the representative ideas as follows:

$$U_T(v) = \frac{\sum_{i=1}^{n} U_T(v_i) \cdot D(v_i, v)^{-2}}{\sum_{i=1}^{n} D(v_i, v)^{-2}} \quad (1)$$

where $v \notin S$ is the idea in question, $U_T(v_i)$ is the utility of a representative idea $v_i$ in *S*, and $D(v_i, v)$ is the Hamming distance between $v_i$ and *v*. The Hamming distance is a measure of dissimilarity between two bit strings, which reflects the number of bits for which two strings vary (Hamming 1950). The true utility function $U_T(v)$ obtained from Equation 1 has a reasonably "smooth" structure in a high-dimensional problem space (i.e., similar ideas tend to have similar utility values, in general). Such a smooth structure of the problem space is necessary for intelligent decision making to outperform unintelligent random trial and error.

Note that the utility landscape construction method described above is different from that of Kauffman's *N-K* fitness landscapes often used in management science (Kauffman 1993, Levinthal 1997, Rivkin 2000). We chose this approach because our method makes it easier and more straightforward to introduce group-level bias, i.e., discrepancy between the true and master utility functions.

Group-level bias is simulated by adding random perturbation when the master utility function $U_M$ is constructed from the true utility function $U_T$, using a similar algorithm as employed by Sayama and Dionne (2015). Specifically, with a bias parameter $\beta$, each bit on representative ideas in *S* is flipped with probability $0.25\beta$ per bit, and a random number within the range $[-\beta, \beta]$ is added to the utility value of each representative idea. Their utility values are then renormalized to the range [0, 1]. The master utility function $U_M$ is generated from these perturbed representative ideas using Equation 1. In this setting, $\beta = 0$ creates a condition with perfect understanding of the problem ($U_M = U_T$) as a collective, while larger values of $\beta$ represents the lack of understanding of the problem.

Moreover, each agent will unconsciously have a different individual utility function, $U_j(v)$ ($j = 1…N$), which is generated by adding random noise to the master utility function $U_M$ so that:

$$U_j(v) \in [\max(U_M(v) - \xi, 0), \min(U_M(v) + \xi, 1)] \quad (2)$$

for all *v*, where $\xi$ is the parameter that determines the variations of utility functions among agents. $\xi = 0$ represents a perfectly homogeneous collective where every agent has exactly the same utility function ($U_i(v) = U_j(v)$ for all *i* and *j*), while larger values of $\xi$ represents a heterogeneous collective made of diverse agents with very different individual utility functions. Figure 1 shows an example of such individual utility functions in contrast to the master utility function. Misunderstanding of the problem by the individual is evident in that the perturbed individual utility function (gray dots) maintains some structures of the master utility function (black dots), but they are not exactly the same. As bounded rational actors, agents are not aware of the full set of alternatives available to them, nor can agents fully specify potential action-potential outcome causal linkages (Gavetti and Levinthal 2000). Therefore agents in our model are not aware of the entire structure of their own individual utility functions. They cannot tell what ideas would produce global maximum/minimum utility values, though they can retrieve a utility value from the function when a specific idea is given, which is a common assumption made in complex global optimization problems (Horst et al. 2000).

**FIGURE 1**

**Master and Individual Utility Functions**

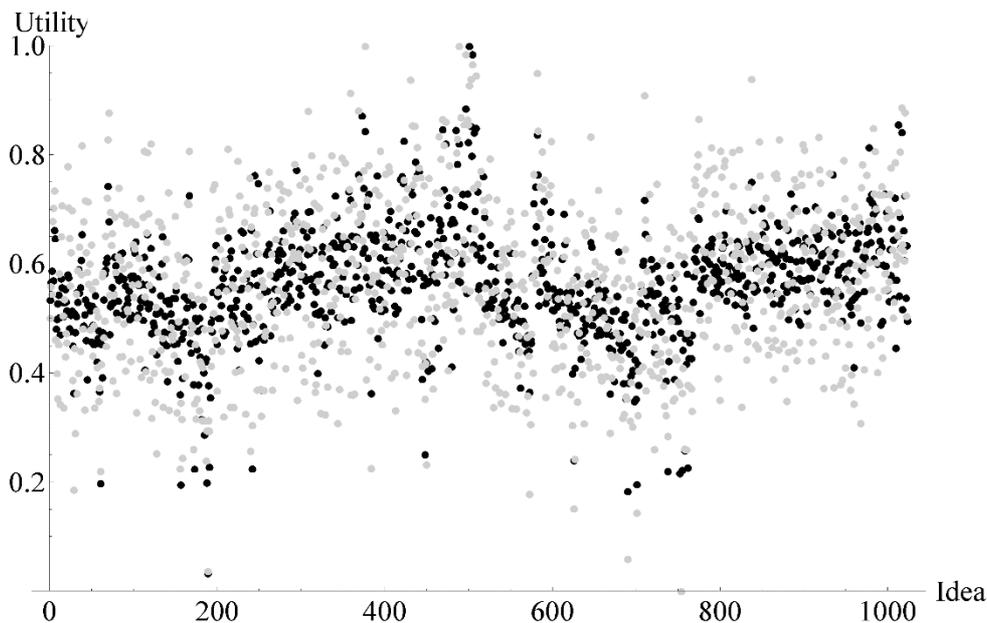

Note: The master utility function with $M = 10$, generated from a representative set of idea utilities of size $n = 10$, is shown by black dots. An individual utility function by adding noise with $\xi = 0.2$ is shown by gray dots. The *x*-axis shows idea indices generated by interpreting bit strings as binary notations of an integer, i.e., all of different ideas are lined up along the horizontal axis and their utility values are plotted.

We recognize that a homogeneous group with no group-level bias would be unlikely in actual groups and collectives. In reality, reduction of a group-level bias would be facilitated by different perspectives, expertise and experiences (i.e., diversity). While varying diversity on any number of dimensions (e.g., ethnic, gender, functional background, education, age) within teams has been studied in the literature (c.f., Kooij-de Bode et al. 2008, O'Reilly et al. 1998, Pelled et al. 1999), research related to group performance has mixed reviews regarding the benefit of diversity within teams. While some diversity is thought to produce a more productive, functional conflict as opposed to an unproductive, relationship conflict (Jehn et al. 1999), a meta-analysis on conflict (De Dreu and Weingart 2003) underscores that these various forms of conflict are all negatively related to group performance. Thus, group-level bias is included in our model to assess potential issues associated with homogeneity within groups.

*Evolutionary operators.* Our model uses agent behaviors reflecting either selection or variation as analogues for decision making behavior: replication, random point mutation, intelligent point mutation, recombination, and subtractive selection. While these five operators reflect common forms of action in evolution (except the intelligent point mutation that does not exist in real biological evolution), they also align with actions commonly found in brainstorming and normative decision making idea generation phases where the goal is to build new ideas from individually generated suggestions (Paulus and Yang 2000) (i.e. mutations and recombination) and idea evaluation phases where culling or supporting ideas (i.e., replication and/or subtraction) leads to final group idea selection and decision. Among those evolutionary operators, replication and subtractive selection use a preferential random search algorithm (Solis and Wets 1981), by which an agent randomly samples $r_p$ ideas from the idea population in its mind, and then the agent selects the best (or worst) idea among the sampled ones for replication (or subtractive selection). Note that the designs of the evolutionary operators used in this model are different from those used in earlier models (Sayama and Dionne 2015), in order to make the variation and selection mechanisms more clearly separable. They are also extended so that their outcomes affect not only the agent's own idea population but also those of its local neighbors on a social network, which represents the exchange of ideas through social ties. In other words, other agents can "hear" the focal agent's opinion and update their own idea population according to it.

Of the five evolutionary operators, replication and subtractive selection are selection-oriented operators, driving the exploitation in the discussion and decision making process. The other three processes (random/intelligent point mutations and recombination) are variation-oriented operators that increase the idea diversity and explore the problem space further. To systematically control and sweep the balance between the two evolutionary "forces" (selection/exploitation and variation/exploration), we introduced a global parameter $p$, which determines the behavioral tendency of agents. Specifically, each agent chooses an exploitative operator with probability $p$ (or, an explorative operator with probability $1 - p$; see Table 2). Setting $p = 1$ makes the agents completely selection-oriented, while $p = 0$ makes them fully exploratory.

*Simulation settings.* Table 2 summarizes the parameter values used in our computer simulations. Most of those values were taken from earlier work (Sayama and Dionne 2015), and were chosen so as to be reasonable in view of typical real collective decision making settings. We tested several variations of parameter settings and confirmed that the results were not substantially different from the ones reported below in this paper.

There are several experimental parameters that we varied in the three sets of computational experiments presented below. The first set of computational experiments manipulated $\beta$, group-level bias, and $\xi$, within-group noise. These two parameters were varied to represent different levels of accuracy and consistency of individual utility functions within a group. The second set of computational experiments varied $p$, the parameter that determines the balance between selection-oriented and variation-oriented operators in agents' behaviors. The third set of computational experiments varied the size and topology of the group, by exponentially increasing the number of agents from $N = 5$, a small group whose size is within the optimal range for decision making teams (Kerr and Tindale 2004, Salas et al. 2004), to $N = 640$, which forms a non-trivial social network. In all cases, the average node degree (i.e., average number of connections attached to a node) was always kept to four, which is a typical number of people one could have meaningful conversations with simultaneously. This assumption made the $N = 5$ case a fully connected network, while the network became increasingly sparse as $N$ increased. For each specific value of $N$, three different network topologies were tested: random (RD), small-world (SW) (Watts and Strogatz 1998) and scale-free (SF) (Barabási and Albert 1999). For small-world networks, the link rewiring probability was set to 10%, which realizes the small-world property (Watts and Strogatz 1998) for relatively small-sized networks like those used in this study. These topological variations do not cause any effective differences for smaller $N$, but as $N$ increases, their influences on network topology and dynamics of idea evolution begin to differentiate.

**TABLE 2**

**Parameters and Symbols**

| Parameter | Value | Meaning |
|---|---|---|
| *Parameters Related to Evolutionary Decision Process* | | |
| $M$ | 10 | Problem space dimensionality |
| $n$ | 10 | Number of representative ideas to generate true/master utility functions |
| $r_p$ | 5 | Number of sample ideas in preferential search algorithm |
| $r_m$ | 5 | Number of offspring generated in intelligent point mutation |
| $p_m$ | 0.2 | Random mutation rate per bit |
| $p_s$ | 0.4 | Probability of random switching in recombination |
| **$p$** | **0~1** | Probability for an agent to take selection-oriented actions |
| | $p/2$ | Probability of replication - advocacy |
| | $p/2$ | Probability of subtractive selection - criticism |
| | $(1-p)/3$ | Probability of random point mutation - minor modification of idea |
| | $(1-p)/3$ | Probability of intelligent point mutation - improvement of existing idea |
| | $(1-p)/3$ | Probability of recombination - generating new ideas from crossing multiple existing ideas |
| *Parameters Related to Team Characteristics* | | |
| $N$ | **5~640** | Size of group/social network |
| **Network topology** | **RD, SW, SF** | RD: random network, SW: small-world network, SF: scale-free network |
| $d$ | 4 | Average degree (average number of links connected to each agent) |
| $k$ | 5 | Number of initial randomly generated ideas in each agent's mind |
| $\beta$ | **0~1** | Group-level bias |
| $\xi$ | **0~1** | Within-group noise |
| $T$ | 60 | Number of iterations |

Note: Bold indicates experimental parameters varied

*Metrics of group performance.* Performance of a group is likely a multidimensional construct, as different authors have tested differing dimensions of group-based adaptation (c.f., LePine 2005, Kozlowski et al. 1999). For the purposes of collective decision making in organizational settings, the ability to converge on a decision is critical, as a group that cannot produce a decision likely fails in their task. In the meantime, convergence on a poor decision may be equally detrimental to a group as well, as mistakes could be costly. As such, it would seem that minimally the consideration of both convergence and decision quality would be needed to assess group performance. As required by increasingly complex organizational environments, groups and organizations need to converge quickly on decisions, and yet ensure these decisions have high efficacy related to solving perceived problems.

We therefore used the two separate performance metrics originally proposed by Sayama and Dionne (2015): one was the true utility value of the mode idea (the most supported idea) in the final population of ideas collected from all the agents' minds, to measure the overall quality of collective decisions. This was selected as it is most likely that the most supported idea represents the group's preferred idea, and once selected, this supported idea will be tested in the context of real-world problem solving.

The other performance metric was the diversity of ideas remaining in the final population of ideas collected from all the agents' minds, to measure the failure of the group to converge. This measurement is based on the classical definition of Shannon's information entropy (Shannon 1948),

$$H = -\sum_{i=1}^{m} p(x_i) \log_2 p(x_i), \qquad (3)$$

where $m$ represents the number of different types of ideas in the final idea population, and $p(x_i)$ is the ratio of the number of the $i$-th type of idea to the total size of the final idea population. The theoretical maximum of $H$ would be $M$, which occurs when all of $2^M$ possible distinct ideas are equally represented. $H$ decreases as the idea population becomes more homogeneous, and it reaches the theoretical minimum 0 when the idea population is made of only instances of the same idea (which would never occur in simulations). To rescale this quantity to the range between 0 and 1, we used $(M - H) / M$ as a measurement of the convergence of final collective decision.

## Results

In this section, we describe our simulation results in three parts: (1) effects of within-group noise and group-level bias (diversity of problem understanding), (2) effects of balance between selection-oriented and variation-oriented behaviors (diversity of behaviors), and (3) effects of group size and social network topology. The first part directly addresses the knowledge/opinion diversity and multi-level issues of collective decision making. The second part illustrates the implications of behavioral diversity of groups for their collective decision performance. Finally, the third part extends our understanding to large-scale, networked organizational settings.

**Part 1: Effects of Within-group Noise and Group-level Bias**

We first conducted a computational experiment to examine the effects of increasing (a) within-group noise, $\xi$, i.e., heterogeneity of individual utility functions within a group, and (b) group-level bias, $\beta$, i.e., discrepancy of the master utility function from the true utility function at a group level, on the overall group performance. For this initial computational experiment, the group was made of five agents with fully connected social network structure (i.e., everyone could talk to everyone else; a small group setting). We assumed that the agents were balanced in terms of their tendency between selection-oriented and variation-oriented behaviors in the discussion (i.e., $p = 1/2$).

Figure 2 presents a summary of the results of simulations with within-group noise $\xi$ and group-level bias $\beta$ systematically varied. Each of the two performance metrics (i.e., level of convergence and utility of most supported idea, as described above) are visualized in a separate 3-D surface plot. We found that the level of convergence was affected significantly by the within-group noise, while it was not affected at all by the group-level bias. On the other hand, the true utility of collective decisions degraded significantly when either the within-group noise or the group-level bias (or both) was increased. The true utility achieved by the most heterogeneous groups ($\xi \sim 1.0$) or the most biased groups ($\beta \sim 1.0$) dropped to about 0.5, which could be achieved just by random idea generation. This means that no net improvement was achieved during the discussion by those groups.

## FIGURE 2

**Effects of Within-Group Noise and Group-Level Bias on Decision Convergence and Quality**

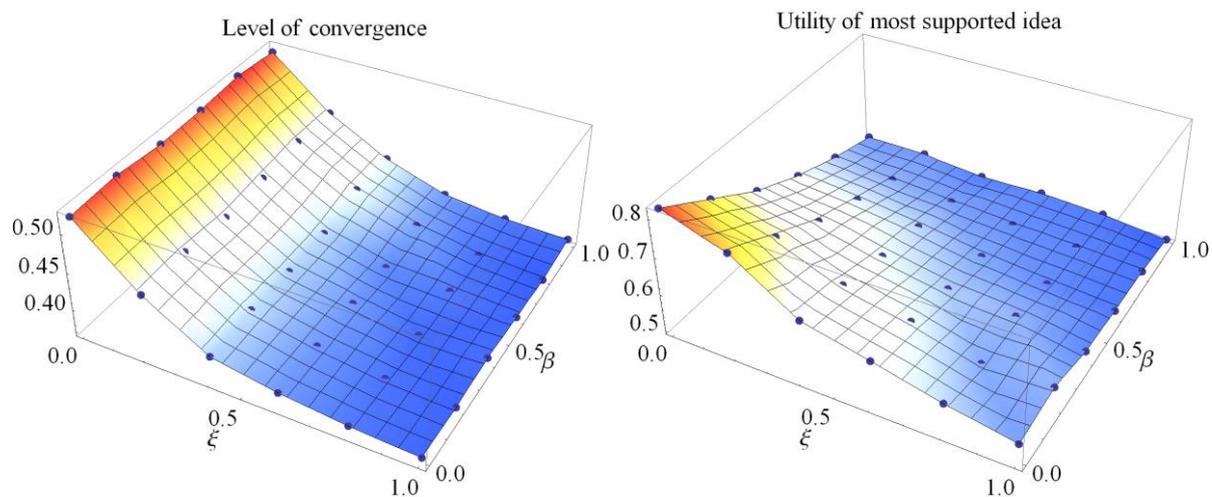

Note: Effects of within-group noise ($\xi$) and group-level bias ($\beta$) on the level of convergence (left) and the true utility value of the most supported idea (right). Each dot represents an average result of 500 independent simulation runs.

**Part 2: Effects of Balance Between Selection-oriented and Variation-oriented Behaviors**

The above computational experiment assumed that the agents' behaviors were well balanced between selection-oriented and variation-oriented operators. We therefore ran another computational experiment to investigate the effects of balance between selection-oriented and variation-oriented behaviors patterns by systematically varying the parameter $p$. Greater values of $p$ represent groups with more selection-oriented behaviors (i.e., advocacy and criticism), while smaller values of $p$ represent groups with more variation-oriented behaviors (i.e., mutations and recombination). The group-level bias, $\beta$, was also varied as another experimental parameter, while the within-group noise, $\xi$, was fixed to 0.2 for this experiment. The group size and their network topology were the same as those in the first computational experiment.

Figure 3 shows a summary of the results of the second computational experiment comparing group performances with different group behaviors, plotting two performance metrics in separate 3-D plots as used for Figure 2 (note that one of the axes is now for $p$, not for $\xi$). The effect of behavioral balance on the level of convergence is straightforward in that greater $p$ (more selection-oriented behaviors) tended to promote convergence more. The effect of *p* on the utility of collective decisions, however, turned out not so trivial. While purely variation-oriented behaviors (*p* ~ 0.0) did not help increase the decision quality, neither did purely selection-oriented behaviors (*p* ~ 1.0). There was a range of optimal balance (*p* = 0.7~0.9) where the groups achieved the highest decision quality. In the meantime, the effect of group-level bias is similar to that seen in Figure 2, so that the utility of collective decisions would be significantly lower if there was group-level bias.

## FIGURE 3

**Effects of Balance between Selection-Oriented and Variation-Oriented Behaviors and Group-Level Bias on Decision Convergence and Decision Quality**

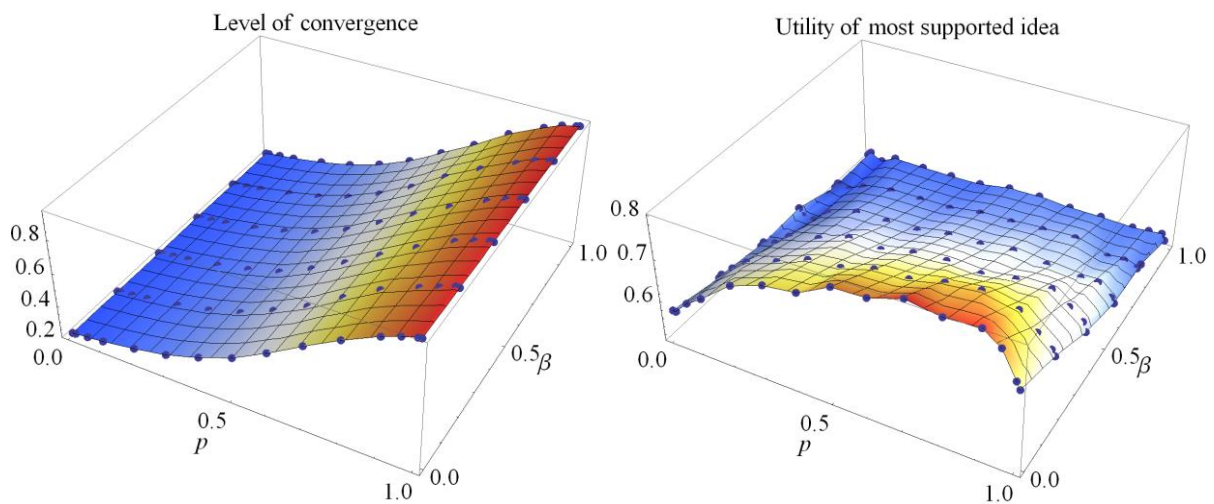

Note: Effects of group-level balance between selection-oriented and variation-oriented behaviors (*p*) and group-level bias ($\beta$) on the level of convergence (left) and the true utility value of the most supported idea (right). Each dot represents an average result of 500 independent simulation runs.

**Part 3: Effects of Group Size and Social Network Topology**

The first computational experiment above assumed small, fully connected networks of agents. While the results produced useful implications for collective decision making in small group settings, they were not sufficient to generate insight into more general collective decision making dynamics on a larger non-trivial social environment, such as in a complex organization or on social media. We therefore conducted the third computational experiment in which the size of groups was increased from 5 to 640 in an exponential manner. For each size of the groups/networks, the average number of connections per agent (i.e., "degree" in network science terminology) were always kept to four, which was the same value as in the first two experiments above. The following values were used for other parameters: $\beta = 0.0$, $\xi = 0.2$, *p* = 0.5.

In this computational experiment, larger groups were no longer considered a typical "group", but rather they formed a more complex social/organizational network, perhaps more indicative of a "collective" in the organizational sciences. For each network size, we used the following three social network topologies. A new network topology was generated for each independent simulation run:

- *Random network* (RD): A random network is a network in which connections are randomly assigned, which can be used as a random control condition. For our computational experiment, a total of $2N$ links were established between randomly selected pairs of agents.
- *Small-world network* (SW) (Watts and Strogatz 1998): A small-world network is a locally clustered (pseudo-)regular network, with a small number of global links introduced to reduce the effective diameter of the network significantly (i.e., a "small-world" effect). The small-world network may be considered a spatially extended network made of mostly local connections but with a few global connections. For our computational experiment, $N$ agents were first arranged in a circle and each agent was connected to its nearest and second nearest neighbors so that the degree would be four for all. Then 10% of the links were randomly selected and either the origin or destination of each of those links was rewired to a randomly selected agent.
- *Scale-free network* (SF) (Barabási and Albert 1999): A scale-free network is a network in which the distribution of node degrees shows a power-law distribution. It represents a heterogeneous network made of a large number of poorly connected nodes and a few heavily connected "hubs". Many real-world networks, including biological, engineered and social networks, were shown to be scale-free (Barabási 2009). While such networks show a small effective diameter like small-world networks, they may not have high local clustering. For our experiment, a well-known preferential attachment algorithm (Barabási and Albert 1999) was used, starting with a fully connected network of five agents and then incrementally adding an agent by connecting it with two links to two existing agents selected preferentially based on their degrees, until the network size reached $N$.

Figure 4 shows the effects of size and topology of networks on the decision outcomes. The larger the group (or network) becomes, the harder it achieves convergence. Apparently there was no substantial difference between the three topological structures regarding their effects on the level of convergence. On the other hand, increasing group size had positive effects on the utility of the most supported idea within the group or on the social network.

**FIGURE 4**

**Effects of Group Size and Social Network Topology on Decision Convergence and Decision Quality**

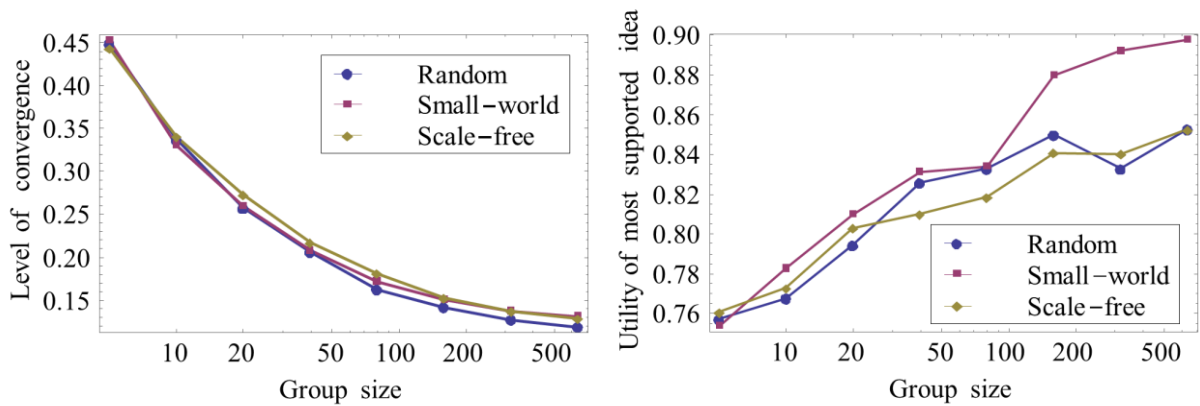

Note: Effects of group size (*N*) and social network topology (random, small-world or scale-free) on the level of convergence (left) and the true utility value of the most supported idea (right). Note the log scale for group size. Each dot represents an average result of 500 independent simulation runs.

One particularly interesting phenomenon seen in Figure 4 is the difference in the utility of collective decisions between small-world networks and other two networks for larger *N* ($N > 100$). Figure 5 provides a more detailed view into this finding, showing the distributions of utilities of most supported ideas for 500 independent simulation runs for $N = 640$ under each of the three conditions. In each condition, the agents were able to find the best idea with utility 1.0 most of the time, but small-world networks facilitated such optimal decision making most frequently. The Mann-Whitney *U* test detected statistically significant differences between small-world and random ($p < 0.003$) as well as small-world and scale-free ($p < 10^{-6}$) networks, while there was no significant difference between random and scale-free ($p = 0.107$) networks. The key distinctive feature of small-world networks that are not present in either random or scale-free networks is the local clustering.

### FIGURE 5

**Distributions of Utilities of Most Supported Ideas Between Different Social Network Topologies**

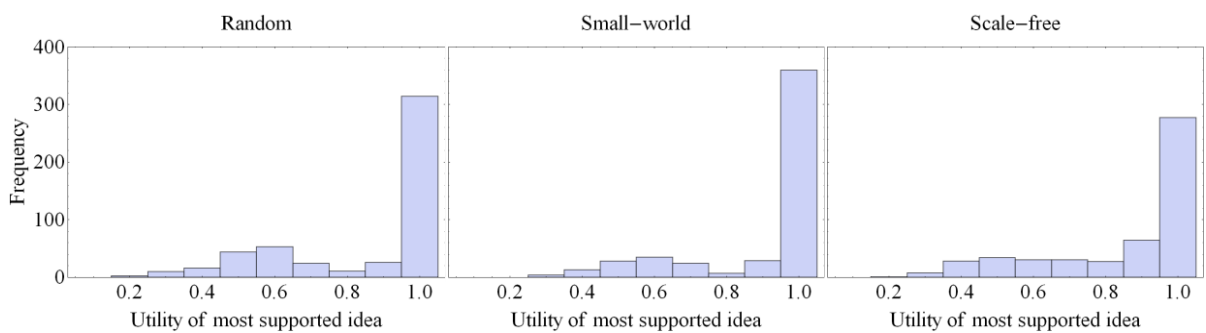

Note: Simulation results comparing the distributions of utilities of most supported ideas at the end of simulation between the three social network topologies (random, small-world or scale-free) for $N = 640$. The small-world network topology (middle) achieved the highest number of the maximal utility value (1.0) compared to the other two topologies, random (left) and scale-free (right).

## Discussion

In this study, we developed an agent-based model and applied evolutionary operators as a means of illustrating how individuals, groups and collectives may move through a decision process based on ecologies of ideas over a social network habitat. We also considered various compositions of group members ranging from homogeneity to heterogeneity and examined the impact of group behaviors on the dynamic decision process as well. These explorations move toward a more realistic view of collective decision making within complex social systems, and answer calls (e.g., Meyer et al. 2005) for research that considers the impact of time and situations in flux, along with nonlinear, multi-level concepts incorporating evolutionary conceptual development. In what follows, we discuss our findings and their implications for human collective decision making.

**On Diversity of Problem Understanding and Multilevel Issues**

Our exploration revealed that the composition of the team or group has implications for decision making and likely considers the complex nature of asking several individuals to come together and agree on a direction that is best suited for the group/collective, rather than for each individual. Research on group diversity has found mixed results related to diversity and group performance issues such as creativity and decision effectiveness (De Dreu and West 2001, Harrison et al. 2002, Hoffman 1979, Jehn and Mannix 2001, Nemeth 1986, 1992, O'Reilly et al. 1989, Wei et al. 2015). Our research, however, indicates an important trade-off between reduction of within-group conflicts and mitigation of group-level bias, as they are not independent from each other. Specifically, if a group is assembled by gathering similar individuals with similar backgrounds, expertise and opinions, then the group tends to have less within-group conflicts but may risk of having a greater group-level bias. On the other hand, if a group is made of diverse individuals with different backgrounds, expertise and opinions, the group may have greater within-group conflicts but it may successfully reduce potential group-level bias and accomplish deeper discussion and better integration of ideas, as the diverse perspectives may represent the actual nature of the problem more correctly.

This means that what kind of strategies of group formation will be optimal to maximize the true utility of collective decisions remains a non-trivial and problem-dependent question, and the best team or group composition may depend greatly on specific problem settings. For example, if a team is tasked to work on a time-critical mission, then the convergence speed is key to their success and thus the emphasis should be placed more on the group homogeneity to avoid within-group conflicts. Or, if a team is formed to seek a truly high-quality solution to a problem, then minimizing the possibility of group-level bias is critical for the team's success, which may require increasing within-group diversity.

**On Diversity of Behaviors and Evolutionary Tendencies**

Our results also imply that the balance between selection-oriented and variation-oriented behaviors may play an important role in collective decision making. Exploration of such behavioral balances was a meaningful step of research because, in realistic organizational settings, some groups may be more prone to be critical, trying to purge bad ideas, while other groups may tend to promote combinations of multiple ideas in discussion. Examples of such behavioral patterns include organizational "cultures" shared by all group members, which is a plausible view of a factor that may influence group dynamics (Salas et al. 2004).

Our results showed that selection-oriented behaviors greatly promoted convergence, yet they were not sufficient to achieve the highest possible utility. To improve the decision quality, the group needs a good mixture of exploratory (variation-oriented) and exploitative (selection-oriented) behaviors. This also ties back to the diversity issue discussed above; a group may not necessarily benefit from diversity of individual problem understanding, but it can benefit from behavioral diversity of group members. In our simulations, the optimal balance between selection and variation was attained at $p \sim 0.8$ (i.e., 80% selection, 20% variation) but this particular balancing point may be problem dependent.

**On Group Size and Social Network Structure**

Finally, our results with social network structure illustrated intriguing effects of group size and network topologies on decision quality, which were manifested particularly for larger networks. Without surprise, the larger the group (or network) becomes, the more elusive convergence on a decision becomes as well. However, group size did positively affect the utility of the most supported idea because, in a large network, agents can conduct different threads of discussions in parallel, which increases the chance for them to collectively find a better idea in the complex problem space. It is important for the agents to remain connected to each other so that the better ideas gradually spread over the network and are widely accepted to become the more supported ideas. The same number of disconnected (non-collaborative, non-interdependent) agents would not be able to achieve this kind of information aggregation and selection task.

A more intriguing finding was obtained regarding the effects of non-trivial network topology. While network topology did not seem to affect level of convergence, small-world networks with spatially localized clusters tended to promote collective search of optimal ideas more often than random or scale-free networks, despite that the network size and the average degree were all identical. Such locally clustered social network structure helps agents in different regions in a network maintain their respective focus areas and engage in different local search, possibly enhancing the effective parallelism of collective decision making and therefore resulting in a greater number of successful decisions. In contrast, random and scale-free networks lack such local clustering, and the links in those networks are all "global", mixing discussions prematurely and therefore reducing the effective parallelism of collective decision making. These observations have an interesting contrast with the fact that random and scale-free networks are highly efficient in information dissemination because of their global connectedness. Our results indicate that such efficiency of information dissemination may not necessarily imply the same for effective collective decision making.

This finding offers another implication for the diversity in collective decision making: certain organizational structures may be more effective in generating and maintaining *idea diversity* in discussion, while other structures would tend to reduce idea diversity and promote premature convergence on suboptimal ideas more often. This is similar to the biological fact that certain geographical habitat structures can maintain greater biodiversity in evolutionary ecology. In the decision making context, this implies that not only within-group diversity or behavioral balance but also social network topologies could influence the dynamics of idea evolution in collective decision making processes.

## Conclusions

In this work, we demonstrated that, using an evolutionary framework to model human collective decision making processes, one can specifically examine the efficacy of a variety of decision processes employed by groups and collectives. The framework we proposed enables a means for direct comparison of various idea evolution paths within collective decision making, and enables an exploration of how the make-up and structure of teams could be critical depending on the overall requirements for decision making tasks. Furthermore, the evolutionary framework and subsequent computational model enables advancements in understanding collective decision making within a dynamic and complex social system. By employing an evolutionary framework we can explore the impact of time and situations in flux, and the modeling enables nonlinear exploration of processes. Finally, the multi-level, network-oriented nature of this research more appropriately models the potential differences in team composition and organizational topologies. It adds to our understanding of the complex nature of collective decisions, and the potential pitfalls and caveats of employing various decision processes and designing teams in a heterogeneous and/or homogeneous manner.

**Limitations and Future Directions**

There are several limitations to our computational modeling study. For example, genetic operators may not exist in groups as "cleanly" as modeled in our simulation. We used simple parameterized settings to control the prevalence of operators, which may not be appropriate to represent the real individual behavior in discussion. Also, our model considered only the heterogeneity of the utility functions of agents. To conduct a more comprehensive, systematic investigation of the homogeneity/heterogeneity issues, it would be critical to incorporate the heterogeneity of the participants' domains of expertise, in addition to their utility functions. Furthermore, we tested only three typical social network topologies, but they are by no means an exhaustive list of possible organizational structures. Conducting computational and human-subject experiments on more realistic social network topologies would add more realistic dynamics to the results, which are among our future research plan.

## Data Availability

The computational simulation data used to support the findings of this study are available from the corresponding author upon request.

## Conflicts of Interest

The authors declare that there is no conflict of interest regarding the publication of this paper.

## Funding Statement

This material is based upon work supported by the National Science Foundation under Grant No. 1734147.